\documentclass[useAMS]{mn2e}
\usepackage{times}
\usepackage{graphicx}

\pubyear{2005}

\author[Bagla, Prasad and Ray]
{J.S.Bagla, Jayanti Prasad and Suryadeep Ray \\
  Harish-Chandra Research Institute,  Chhatnag Road, Jhusi, Allahabad 211019,
  India. \\ 
  E-mail: (jasjeet, jayanti, surya)@mri.ernet.in} 

\title[Gravitational collapse and the role of substructure]
{Gravitational collapse in an expanding background and the role of
  substructure I: Planar collapse}

\label{firstpage}

\def\LaTeX{L\kern-.36em\raise.3ex\hbox{a}\kern-.15em
    T\kern-.1667em\lower.7ex\hbox{E}\kern-.125emX}

\pagerange{\pageref{firstpage}--\pageref{lastpage}} \pubyear{2004}

\begin{document}

\maketitle


\begin{abstract}
We study the interplay of clumping at small scales with the collapse
and relaxation of perturbations at much larger scales.  
We present results of our analysis when the large scale perturbation is
modelled as a plane wave.  
We find that in absence of substructure, collapse leads to formation of a
pancake with multi-stream regions.  
Dynamical relaxation of plane wave is faster in presence of substructure.
Scattering of substructures and the resulting enhancement of transverse
motions of haloes in the multi-stream region lead to a thinner pancake.
In turn, collapse of the plane wave leads to formation of more massive
collapsed haloes as compared to the collapse of substructure in absence of the
plane wave.   
The formation of more massive haloes happens without any increase in
the total mass in collapsed haloes.  
A comparison with the Burgers' equation approach in absence of any
substructure suggests that the preferred value of effective viscosity depends
primarily on the number of streams in a region.  
\end{abstract}




\section{Introduction}

Large scale structures like galaxies and clusters of galaxies are
believed to have formed by gravitational amplification of small
perturbations \cite{lssu,peacock,tpvol3,lss_review}.  
Observations suggest that the initial density perturbations were present at
all scales that have been probed by observations.  
An essential part of the study of formation of galaxies and other large scale
structures is thus the evolution of density perturbations for such initial
conditions.  
Once the amplitude of perturbations at any scale becomes large, i.e., $\delta
\sim 1$, the perturbation becomes non-linear and the coupling with
perturbations at other scales cannot be ignored.  
Indeed, understanding the interplay of density perturbations at different
scales is essential for developing a full understanding of gravitational
collapse in an expanding universe. 
The basic equations for this have been known for a long time \cite{deleqn} but
apart from some special cases, few solutions are known.  

A statistical approach to this problem based on pair conservation
equation has yielded interesting results \cite{hamil,rntp,tp96,ekp2000}, and
these results have motivated detailed studies to obtain fitting functions to
express the non-linear correlation function or power spectrum in terms of the
linearly evolved correlation function \cite{hamil,jmw,pd96,st_clus_n}.

It is well known from simulation studies that at the level of second
moment, i.e., power spectrum, correlation function, etc., large scales have an
important effect on small scales but small scales do not have a significant
effect on large scales \cite{renorm1,klmel,litwein,trpwr,renorm2}.  
Most of these studies used power spectrum as the measure of clustering.  
Results of these simulation studies form the basis for the use of N-Body
simulations, e.g., from the above results we can safely assume that small
scales not resolved in simulations do not effect power spectrum at large
scales and can be ignored.  

Substructure can play an important role in the relaxation process.  
It can induce mixing in phase space \cite{violent,noise}, or
change halo profiles by introducing transverse motions
\cite{previr,ks_halo}, and, gravitational interactions between
small clumps can bring in an effective collisionality even for a collisionless 
fluid \cite{ma,kin_halo}. 
Thus it is important to understand the role played by substructure in
gravitational collapse and relaxation in the context of an expanding
background. 
In particular, we would like to know if this leaves an imprint on the
non-linear evolution of correlation function.  
Effect of substructure on collapse and relaxation of larger scales is another
manifestation of mode coupling.

In this paper, we report results from a study of mode coupling in
gravitational collapse.  
In particular, we study how presence of density perturbations at small
scales influences collapse and relaxation of perturbations at larger scales. 
These effects have been studied in past \cite{previr_sim} but the
motivation was slightly different \cite{previr}.  
We believe it is important to study the issue in greater detail and make the
relevance of these effects more quantitative using N-Body simulations with a
larger number of particles.  
We also study the reverse process, i.e., how does collapse of perturbations at
large scales effect density perturbations at much smaller scales. 

It is well known that the local geometry of collapse at the time of initial
shell crossing is planar in nature \cite{za}, hence we model density
perturbations as a single plane wave in this work.  
Simple nature of the large scale fluctuation allows us to study interaction of
well separated scales without resorting to statistical estimators like power
spectrum.  
We are studying the same problem in a more general setting and those results
will be reported in a later publication.   

Key features of collapse of a plane wave can be understood using quasi-linear
approximations, at least at a qualitative level.
Initial collapsing phase is well modelled by the Zel'dovich approximation
\cite{za}, wherein particles fall in towards the centre of the potential
well. 
Zel'dovich approximation breaks down after orbit crossing as it does not
predict any change in the direction of motion for particles, thus in this
approximation particles continue to move in the same direction and the size of
the collapsed region grows monotonically.  
In a realistic situation we expect particles to fall back towards the
potential well and oscillate about it with a decreasing amplitude, and the
collapsed region remains fairly compact. 
Several approximations have been suggested to improve upon the Zel'dovich
approximation
\cite{adh,1989RvMP...61..185S,adh2,ffa,lep,fpa,approxrev1,approxrev2}. 
The adhesion approximation \cite{adh,adh2} invokes an effective
viscosity: this prevents orbit crossing and conserves momentum to
ensure that pancakes remain thin and matter ends up in the correct region. 
This changes the character of motions in dense regions (no orbit crossing or
mixing in the phase space) but predicts locations of these regions correctly.  
If one assumes that the gravitational potential evolves at the linear rate
\cite{lep,fpa}, then it can be shown that the collapsed region remains
confined.
The effective drag due to expanding background slows down particles and they
do not have enough energy to climb out of the potential well.  

Thus the process of confining particles to a compact collapsed region results
from a combination of expansion of the universe and gravitational interaction
of in falling particles. 
None of the approximations captures all the relevant effects.
Therefore we must turn to N-Body simulations \cite{bertsc98,nbreview} in order
to study collapse and relaxation of perturbations in a complete manner.


\section{Evolution of Perturbations}

We will consider only gravitational effects here and ignore all other
processes. 
We assume that the system can be described in the Newtonian limit. 
The growth of perturbations is then described by the coupled system of Euler's
equation and Poisson equation in comoving coordinates along with mass
conservation, e.g., see \cite{lssu}.  
\begin{eqnarray}
  {\bf \ddot x}_i +  2 \frac{\dot{a}}{a} {\bf \dot x}_i 
  &=& - \frac{1}{a^2} \nabla_i \varphi          \nonumber \\
\nabla^2\varphi = 4 \pi G a^2 \left(\rho - \bar\rho\right) &=&    \frac{3}{2}
  \frac{\delta}{a^3} H_0^2 \Omega_{nr}    \nonumber \\ 
\rho({\bf x}) &=& \frac{1}{a^3} \sum\limits_i m_i \delta_{_D}({\bf x} - {\bf x}_i) 
\end{eqnarray}
It is assumed that the density field is generated by a distribution of
particles, each of mass $m_i$, position ${\bf x}_i$.
$H_0$ is the present value of Hubble constant, $\Omega_{nr}$ is the present
density parameter for non-relativistic matter and $a$ is the scale factor. 
In this paper we will consider the Einstein-de Sitter universe as the
background, i.e., $\Omega_{nr}=1$.
These can be reduced to a single non-linear differential equation for density
contrast \cite{deleqn}. 
\begin{equation}
\ddot\delta_{\bf k} + 2 \frac{\dot a}{a} \dot\delta_{\bf k} =
\frac{3}{2} \frac{\delta_{\bf k}}{a^3} H_0^2  + A - B
\label{deltak}
\end{equation}
where
\begin{equation}
A = \frac{3}{4} \frac{1}{a^3} H_0^2
\sum\limits_{{\bf k'} \neq 0, {\bf k}} \left[ \frac{{\bf k}.{\bf
k'}}{{k'}^2} + \frac{{\bf k}.\left({\bf k} - {\bf k'}
\right)}{\left|{\bf k} - {\bf k'} \right|^2} \right] \delta_{\bf k'}
\delta_{{\bf k} -{\bf k'}} \nonumber
\end{equation}
and
\begin{equation}
B = \frac{1}{M} \sum\limits_i m_i \left({\bf k}.{\bf
\dot x}_i \right)^2 \exp\left[ i {\bf k}.{\bf x}_i \right] \;\;\;
; \;\;\;\; M= \sum\limits_i m_i \nonumber
\end{equation}
The terms $A$ and $B$ are the non-linear coupling terms between
different modes. 
$B$ couples density contrasts in an indirect manner through velocities of
particles (${\bf \dot x}_i$). 
The equation of motion still needs to be solved for a complete solution of
this equation, or we can use some ansatz for velocities to make this an
independent equation. 

It can be shown that individual {\it virialised} objects, i.e., objects that
satisfy the condition $2T+U=0$ where $T$ is the kinetic energy and $U$ is the
potential energy, do not make any contribution towards growth of perturbations
through mode coupling \cite{deleqn} at much larger scales, i.e., the $A-B$
term is zero.
The contribution of mode coupling due to interaction of such objects is not
known. 

Approximate approaches to structure formation can be developed by ignoring
interaction of well separated scales.
The evolution of density perturbations can be  modelled as a combination of
non-linear collapse at small scales, and the collapsed objects can be
displaced using quasi-linear approximations
\cite{pp1,pp2,pp3,pinocchio1,pinocchio2,pinocchio3}. 
These approaches yield an acceptable description of properties of collapsed
objects and their distribution for a first estimate. 
PINOCCHIO \cite{pinocchio1,pinocchio2,pinocchio3} provides sufficient
information about halo properties and merger trees for use with
semi-analytic models of galaxy formation. 
The efficacy of these models puts an upper bound on the effects of
mode coupling that we are studying here. 

In this paper we simplify the system by starting with perturbations that have
a non-zero amplitude only for two sets of scales.
We simulate the collapse of a plane wave by starting with non-zero amplitude
of perturbations for the fundamental mode of the simulation box along the $z$
axis, the wavenumber of the fundamental mode is denoted by $k_f$. 
This serves as the large scale perturbation in our study. 
The amplitude for this mode is chosen so that shell crossing takes place when
the scale factor $a=1$. 
Power spectrum for small scale fluctuations is chosen to be non-zero in a
range of wave numbers $k_0 \pm \delta k$ with a constant amplitude across this
window, i.e., $\Delta_s^2(k) = \alpha A$ for $k_0 - \delta k \leq k \leq k_0 +
\delta k$. 
A Gaussian random realisation of this power spectrum is used for small scale
fluctuations. 
Here $\Delta_s^2(k)$ is the power per logarithmic interval in $k$ contributed
by small scales (large $k$) and $A$ is the amplitude of the fundamental mode
that gives rise to the plane wave.
The ratio of $\Delta^2(k)$ at $k=k_0$ and for the plane wave is denoted by
$\alpha$, thus when $\alpha=1$ collapse of perturbations at these scales
happens at the same time whereas for $\alpha > 1$ perturbations at small
scales collapse before the plane wave collapses.
We chose the ratio $k_0/k_f = 8$ so that there is distinct separation in the
scales involved.

\begin{figure}
\includegraphics[width=3.3truein,angle=270]{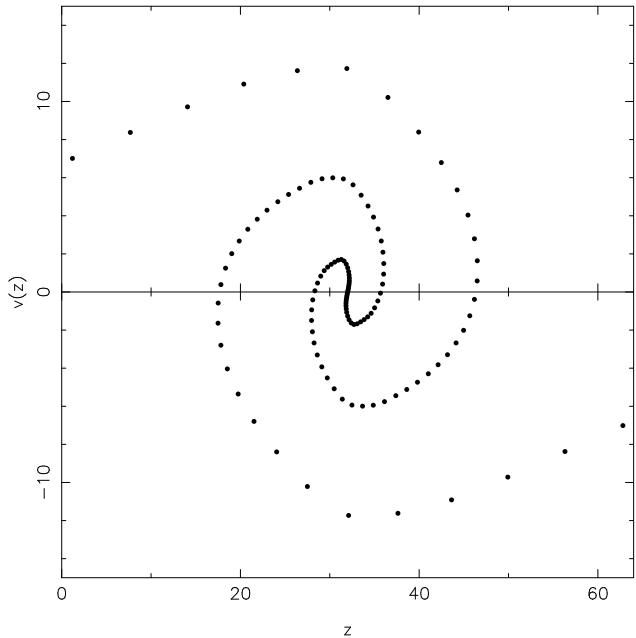}
\caption{Phase space plot for the plane wave at late stages of
  collapse for the simulation PM\_00L.  Velocity of particles is
  plotted as a function of position.  Regions where particles with
  different velocities can be found are the multi-stream regions.  As
  we approach the centre of pancake located at $z=32$, we go from a
  single stream region to a three stream region and so on up to seven
  stream region near the centre.} 
\label{phase_space}
\end{figure}

Collapse of a plane wave of this kind leads to formation of a multi-stream
region, we will also use the term pancake to describe this region.
Figure~\ref{phase_space} shows the phase space plot for the plane wave
at late stages of collapse for the simulation PM\_00L (see table 1 for details
of the simulation).  
Velocity of particles is plotted as a function of position, only the
$z$ component is plotted as there is no displacement or velocity along
other directions in this simulation. 
Regions where particles with different velocities can be found are the
multi-stream regions.  
As we approach the centre of pancake located at $z=32$, we
go from a single stream region to a three stream region and so on up
to seven stream region near the centre. 

In initial stages, the mass in the pancake increases rapidly as more
particles fall in.
Figure~\ref{den_plwave} shows this in terms of over-density which increases
sharply from $a=1$ to $a=2$. 
A significant fraction of the total mass falls into the pancake and
the infall velocities for the remaining matter are very small.  
In this regime the mass of the pancake is almost constant, this can be seen
from the panels of figure~\ref{den_adh_sub} where the mass enclosed in the
pancake region is almost constant from $a=2$ to $a=4$. 

\begin{table}
\caption{This table lists parameters of N-Body simulations we have used.  All
  the simulations used $128^3$ particles.  The first column lists name of the
  simulation, second column lists the code that was used for running the
  simulation, third column gives the relative amplitude of small scale power
  and the plane wave, the fourth column tells us whether the large scale plane
  wave was present in the simulation or not, and the last column lists the
  distribution of 
  particles before these are displaced using a realisation of the power
  spectrum.  {\sl Grid} distribution means that particles started from grid
  points. {\sl Perturbed grid} refers to a distribution where particles are
  randomly displaced from the grid points, this displacement has a maximum
  amplitude of $0.05$ grid points.  Such an initial condition is needed to
  prevent particles from reaching the same position in plane wave collapse as
  such a situation is pathological for tree codes.  The TreePM simulations
  were run with a force softening length equal to the grid length.}
\begin{tabular}{||l|l|l|l|l||}
\hline
Name & Method & $\alpha$ & Plane wave & IC \\
\hline
PM\_$00$L & PM & $0.0$ & Yes & Grid \\
T\_$00$L & TreePM & $0.0$ & Yes & Perturbed grid \\
T\_$05$L & TreePM & $0.5$ & Yes & Grid \\
T\_$10$L & TreePM & $1.0$ & Yes & Grid \\
T\_$20$L & TreePM & $2.0$ & Yes & Grid \\
T\_$40$L & TreePM & $4.0$ & Yes & Grid \\
T\_$10$P & TreePM & $1.0$ & Yes & Perturbed grid \\
T\_$40$P & TreePM & $4.0$ & Yes & Perturbed grid \\
T\_$05$ & TreePM & $0.5$ & No & Grid \\
T\_$10$ & TreePM & $1.0$ & No & Grid \\
T\_$20$ & TreePM & $2.0$ & No & Grid \\
T\_$40$ & TreePM & $4.0$ & No & Grid \\
\hline
\end{tabular}
\end{table}

In absence of any substructure the collisionless collapse retains
planar symmetry and we have layers of multi-stream regions with the
number of streams increasing towards the centre of the pancake.
Presence of small scale fluctuations can induce transverse motions and these
motions are amplified in the pancake.

Weakly bound substructure can be torn apart due to interaction with rapidly
in falling matter.
On the other hand, higher average density in the multi stream region can lead
to rapid growth of perturbations.
It is known that pancakes are unstable to fragmentation due to growth of
perturbations \cite{pancake_2d}.
The velocity field is anisotropic due to infall along one direction, hence the
rate at which perturbations grows will also exhibit anisotropies.
Velocity dispersion along the direction of plane wave collapse is larger than
the transverse direction, hence the growth of fluctuations in the transverse
plane is expected to be more rapid. 

If the in falling material contains collapsed substructure, then gravitational
interactions between these can induce large transverse velocities.
This takes away kinetic energy from the direction of infall, which in turn can
lead to more fragmented and thinner multi-stream region.

In the following sections we describe the numerical experiments have
undertaken in detail, and test the physical ideas and expectations outlined
above. 


\section{Numerical Experiments and Results}

We used a Particle-Mesh code \cite{pm} and the TreePM code
\cite{treepm,treepm_err}. 
Some simulations were run using the parallel TreePM \cite{treepm_par}. 
TreePM simulations used spline softening with softening length equal to the
length of a grid cell in order to ensure collisionless evolution.
We used force softening assuming a spline kernel \cite{gadget}. 
All the simulations were carried out with $128^3$ particles.
Table~1 lists parameters of the simulations we have used for this paper.
We have used two types of initial distribution of particles.  
In the {\sl Grid} distribution particles are located at grid points before
being displaced to set up the initial perturbations.
{\sl Perturbed grid} refers to a distribution where particles are randomly
displaced from the grid points \cite{pm}, this displacement has a
maximum amplitude of $0.05$ grid length.  
Such an initial condition is needed to prevent particles from reaching the
same position in plane wave collapse as such a situation is pathological for
tree codes. 
These small displacements do not affect the power spectrum to be realised,
PM\_$00$L and T\_$00$L were compared to test for any systematic effects.

Simulations T\_$10$P and T\_$40$P were similar to T\_$10$L and
T\_$40$L except that the small scale fluctuations were restricted to
the direction orthogonal to the direction of plane wave. 
Thus the small scale fluctuations had the same form for all $z$.  
These simulations are useful for differentiating between competing
explanations for results outlined below. 

In addition to the N-Body simulations listed in table~1, we also
carried out one dimensional simulations within the adhesion model
\cite{adh} with a finite viscosity following a method similar to the
one outlined by Weinberg and Gunn (1990). 

\begin{figure}
\includegraphics[width=3.3truein]{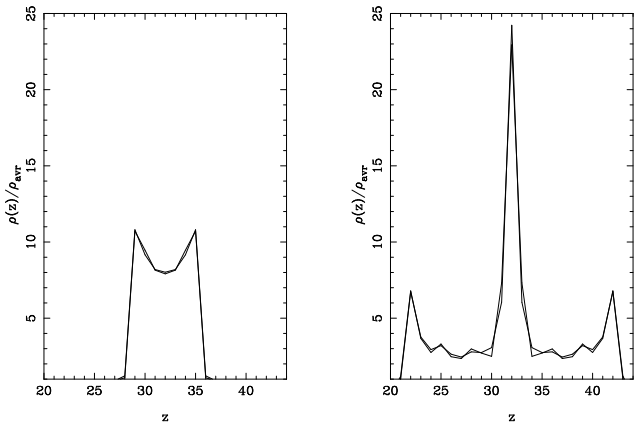}
\caption{Density profile is plotted for two epochs for simulations
  PM\_00L and T\_00L -- in these simulations density varies only in
  the direction along the plane wave.  Solid lines show the density
  profile at $a=1$ and $a=2$ from the PM\_00L simulation, dashed lines
  show the density profile from the T\_00L simulation with the same
  profile.} 
\label{den_plwave}
\end{figure}

Figure~\ref{den_plwave} shows the density profile of the pancake for 
PM\_00L and T\_00L simulations at two epochs.  
These figures demonstrate that the density profiles in these
simulations are almost identical, indeed the tiny differences can be
attributed to the different initial distribution of particles.
We have checked this assertion by running the PM\_00L with the
perturbed grid initial conditions.
The TreePM method has a slightly better resolution but it does not
induce any new features. 
This is expected as the force softening length used in the TreePM
simulations is one grid length, same as the average inter-particle
separation and it has been shown than such force softening does not
induce two body collisions \cite{discr,obliquewave}. 
We will mostly use TreePM simulations for the remaining part of this
study. 

\begin{figure*}
\includegraphics[width=7truein]{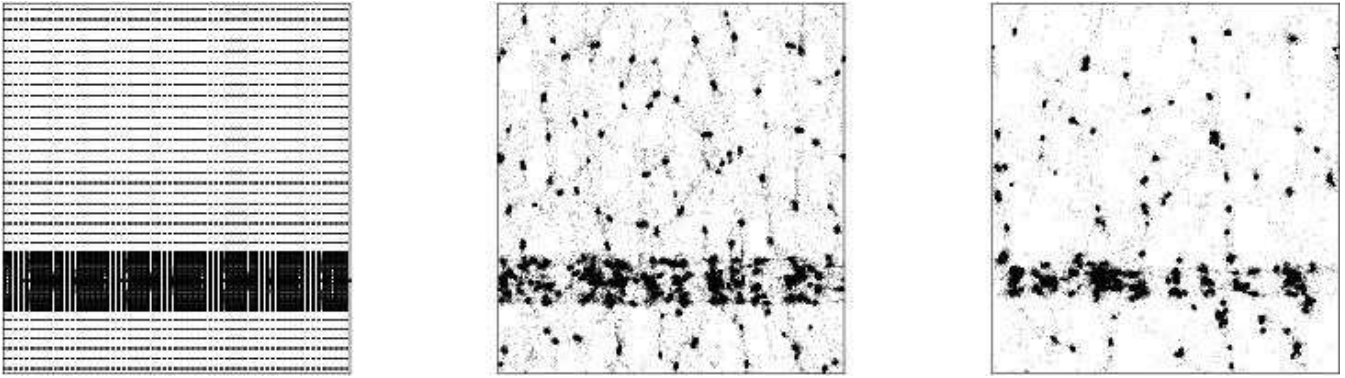}
\caption{Panels in this figure show the same slice from simulation T\_00L,
  T\_10L and T\_40L.  These slices are shown for $a=2$, the plane wave begins
  to collapse at $a=1$.  Plane wave collapses along the vertical direction in
  these slices.  The left panel is for T\_00L, middle panel is for T\_10L and
  the right panel is for T\_40L.}
\label{pthick}
\end{figure*}


\subsection{Thickness of pancake}

An important indicator of the role played by substructure is the
thickness of the pancake that forms by collapse of the plane wave.
If substructure does not play an important role in evolution of
large scale perturbations then the thickness of pancake should not
change by a significant amount.
On the other hand, if substructure does indeed speed up the process of
dynamical relaxation then we should see some signature in terms of the
thickness of pancake, velocity structure, or both.
Any such effect will be apparent only at late times as infall of
matter into the pancake dominates at early times. 
Dynamical effects of substructure will become important only at late
times. 

Figure~\ref{pthick} shows a slice from some of the simulations listed in
table~1. 
The plane wave collapses along the vertical axis.
Configuration at $a=2$ is shown here, the plane wave begins to collapse at
$a=1$. 
Different panels in this figure refer to simulation T\_00L, T\_10L and
T\_40L. 
The boundary of the multi stream region is visible clearly in all the
slices even though this region is fragmented in the last panel (T\_40L). 
It is clear that the pancake is thinner in simulations with more
substructure.   

\begin{figure}
\includegraphics[width=3.3truein]{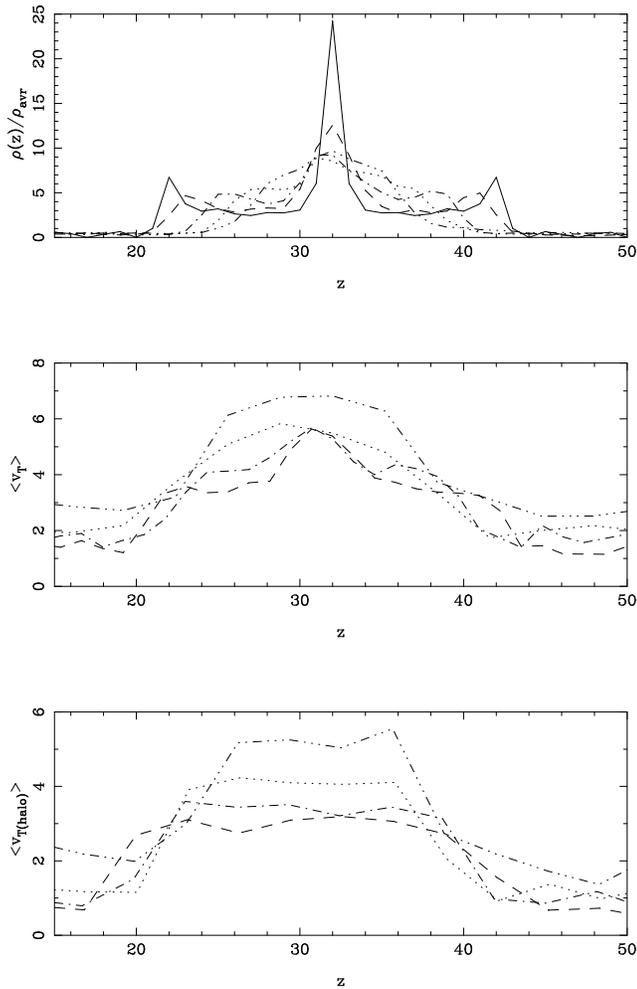}
\caption{Top panel in this figure shows the density profile as a function of
  $z$, the direction of collapse for the plane wave.  Density profile has been
  averaged over the directions transverse to the collapse of plane wave.  The
  curves are for $a=2$, simulations used are T\_00L (solid line), T\_05L
  (dashed line), T\_10L (dot-dashed line), T\_20L (dotted line) and T\_40L
  (dot-dot-dashed line). 
  The middle panel shows the rms transverse velocities of particles at the
  same epoch for T\_05L (dashed line), T\_10L (dot-dashed
  line), T\_20L (dotted line) and T\_40L (dot-dot-dashed line). 
  The lower panel shows the rms transverse velocities of collapsed
  haloes at the same epoch for T\_05L (dashed line), T\_10L (dot-dashed
  line), T\_20L (dotted line) and T\_40L (dot-dot-dashed line).}
\label{den_all}
\end{figure}

A more detailed comparison of simulations with different level of
substructure is shown in figure~\ref{den_all}.
The top panel of this figure shows the averaged over-density as a
function of the $z$ coordinate, the plane wave collapses along this
axis. 
Over-density is averaged over all $x$ and $y$ for a given interval
$(z\pm \Delta z)$ to obtain the averaged values plotted here.  
The peak over-density at the centre of the pancake is smaller in
simulations with more substructure.
The mass enclosed within a given distance of the centre of pancake
(defined here as the trough of the potential well of the plane wave)
is smaller for more substructure, even though the variation is
very small at less than $10\%$ between the extreme cases (see
figure~\ref{den_adh_sub}).  
Potential wells corresponding to substructure prevent infall into the
pancake region. 
As the amount of substructure is increased, there is visible reduction
in the size of the region around the pancake where density is greater
than average. 
The visual impression of figure~\ref{pthick} is reinforced by the
variation of over-density. 

The middle panel of figure~\ref{den_all} shows the {\it rms}
velocities of particles in direction transverse to the plane wave
collapse as a function of the $z$ coordinate.  
As in the top panel, averaging is done over all $x$ and $y$ for a
given interval $(z\pm \Delta z)$.
This plot shows that the transverse motions are enhanced in the dense
pancake region.  
The amplitude of transverse motions is larger in simulations with more
substructure.   
Size of the region where these motions are significant varies with the
amount of substructure, as in case of over-density (top panel). 
The {\it rms} transverse velocities do not go to zero outside the
pancake region, instead these level off to a small residual value.

Transverse motions are due to motions of particles in clumps that
constitute substructure, due to infall of particles in these clumps,
and, transverse motions of clumps as they move towards each other. 
In order to delineate these effects, we have plotted the {\it rms}
velocities for haloes in the last panel of figure~\ref{den_all}. 
These haloes were selected with the friends-of-friends (FOF) algorithm
using a linking length of $l=0.2$ grid length. 
Transverse component of the velocity of centre of mass for haloes with
more than $50$ particles was used for the figure. 
Such a high cutoff for halo members is acceptable because typical
haloes have several hundred members, see the following subsection on
mass functions.  
Differences between simulations with different amount of substructure
are more pronounced than in the middle panel. 
For simulations with a small amount of substructure, motion of clumps
is subdominant and hence the transverse motions are contributed mostly
by internal motions and infall.  
In simulations with more substructure, motions of clumps contribute
significantly to the {\it rms} transverse velocity.
Gravitational attraction of clumps, particularly in close encounters
in the pancake region induce the transverse component.
Collisions are enhanced in the pancake region as the number density of
clumps is higher. 

In order to convince ourselves that transverse motions induced by
scattering/collision of clumps is the most likely reason for the
reduced thickness of pancakes, we compare simulations T\_$10$L and
T\_$40$L with T\_$10$P and T\_$40$P. 
In T\_$10$P and T\_$40$P simulations, the small scale fluctuations do not have
any $z$ dependence.   
In these (T\_$10$P and T\_$40$P) simulations there are no clumps but
streams of particles that are falling in and this reduces the number
of scattering that take place -- no $z$ dependence means that dense
streams run into each other head on with grazing collisions happening
only rarely. 
Of course, in the simulation the presence of the plane wave leads to breaking
of these streams into clumps as the streams are stretched inhomogeneously in
the $z$ direction.  
These clumps are aligned parallel to the $z$ axis. 
In the pancake region scattering of these streams occasionally leads to
complex patterns. 

\begin{figure*}
\includegraphics[width=7truein]{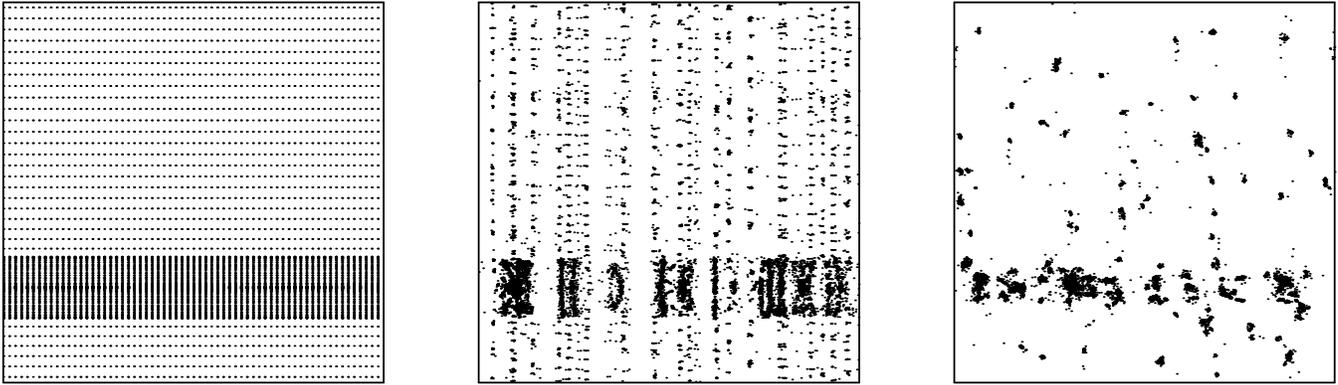}
\caption{This figure shows slices from simulations T\_$40$L and
  T\_$40$P for $a=2$.  The left panels shows a slice from the simulation
  PM\_$00$L, plotted here for reference.  The central panel is for T\_$40$P
  and the right panel is for T\_$40$L.  This visual comparison shows that the 
  pancake is thinner in T\_$40$L as compared to T\_$40$P.  Indeed, the
  thickness of pancake in T\_$40$P and PM\_$00$L is very similar.}
\label{pic_lp}
\end{figure*}

If presence of substructure and its growth in the pancake was the only
cause for making the pancake thinner then pancake in these simulations should
be thinner as well.
Figure~\ref{pic_lp} shows slices from simulations T\_$40$L and
T\_$40$P for $a=2$.  
A slice from the simulation PM\_$00$L is also plotted here for
reference.  
This visual comparison shows that the pancake is thinner in T\_$40$L
as compared to T\_$40$P.  
Indeed, the thickness of pancake in T\_$40$P and PM\_$00$L is very
similar. 
This reinforces the point that scattering of clumps in the pancake
region is the key reason for thinner pancakes.  


\subsection{Pancakes and Viscosity}

The substructure is helping to confine the pancake to a smaller
region.   
It is interesting to study the collapse of a plane wave in an N-Body 
simulation and compare it with the collapse in the adhesion model
\cite{adh} with a finite effective viscosity.
We first study the collapse of a plane wave in absence of any
substructure, N-Body simulations PM\_00L for comparison with numerical
simulations using the adhesion model with finite effective viscosity.  
One dimensional adhesion simulations were done using the plane wave
with the same amplitude as the N-Body simulations.  
We use the standard method for computing the trajectories of particles
in the adhesion model \cite{adh2}, a summary of the basic formalism is
reproduced here for reference.  

In Adhesion approximation, the equation of motion for a particle is
replaced by the Burgers' equation \cite{adh,adh2}.  
In the one dimensional situation we are considering here, we have:
\begin{equation}
\frac{\partial u}{\partial b} + u \frac{\partial u}{\partial x} = \nu
\frac{\partial^2 u}{\partial x^2} .
\end{equation}
Here $u=\partial x / \partial b$ is the 'velocity' of particles and
$b$ is the linear growth factor.  
This equation can be solved by introducing the velocity potential $u =
\partial\psi /\partial x$, where $\psi$ coincides with the gravitational
potential at the initial time.
Solution has the following form.
\begin{equation}
u = \nabla \psi = -2 \nu \nabla \ln U 
\end{equation} 
and, 
 \begin{equation}
U(x,b) = ({1\over { 4 \pi \nu b }})^{1/2} 
\int\limits_{-\infty}^{\infty}
 \exp\left[- \frac{\psi(q)}{2 \nu } - \frac{ \left(x - q\right)^2}{4
     \nu b} \right] dq  . 
\end{equation} 
Here $q$ is the Lagrangian position of the particle and $x$ is the
Eulerian position. 
In this method we integrate the differential equation for particle
trajectories. 
At each time step velocity is calculated by above procedure at grid
points and interpolated to particles positions. 

\begin{figure}
\includegraphics[width=3.3truein]{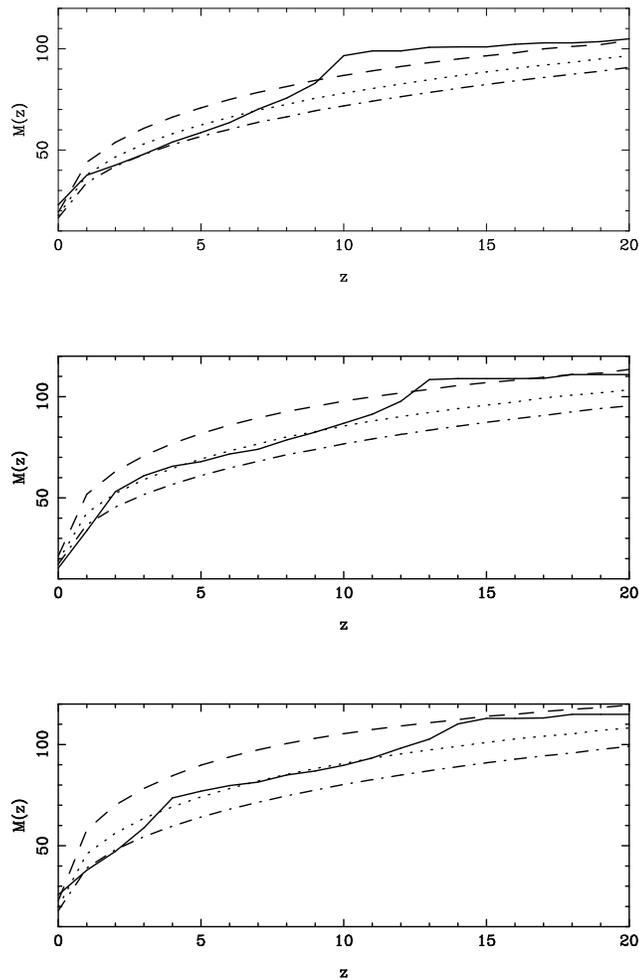}
\caption{This figure shows the mass enclosed within a distance $z$
  from the centre of the multi stream region.  The top panel shows the
  curves for $a=2$.  The thick solid curve is for the N-Body
  simulation PM\_00L.  Jumps in the mass enclosed occur at transition
  from multi stream region with $2n+1$ streams to $2n+3$ streams, with
  $n$ a non-zero positive integer.  All other curves show $M(z)$ for
  adhesion model: dashed curve is for $\nu=400$, dotted curve is for
  $\nu=600$ and the dot-dashed curve is for $\nu=900$.  The middle
  panel shows the same set curves for $a=3$ and the lower panel is for
  $a=4$.} 
\label{den_adh}
\end{figure}

Figure~\ref{den_adh} shows the mass enclosed within a distance $s$ from
the centre of the pancake.  
The enclosed mass is defined as:
\begin{equation}
M(z) = \int\limits_{z_c}^{z_c + z} dz \rho(z+z_c) .
\end{equation}
Here $\rho(z)$ is the density at position $z$ and $z_c$ is the centre of the
pancake. 
There is no ambiguity for comparing the results with N-Body
simulations in case of no substructure as density depends only on
$z$. 
While comparing other simulations with the adhesion solution, we will
consider density averaged over $x$ and $y$ directions -- Adhesion
model is run only for the one dimensional problem.
Top panel of figure~\ref{den_adh} shows the enclosed mass $M(z)$ at
$a=2.0$, middle panel is for $a=3$ and the lower panel is for $a=4.0$.    
The solid curve shows the enclosed mass for PM\_00L.  
In the region with a given number of streams, the N-Body curve is
smooth. 
Jumps in mass enclosed occur at transition from single stream to multi
stream region, and at other transitions where the number of streams
changes within the multi stream region.  
All other curves show $M(z)$ for adhesion model: dashed curve is for
$\nu=400$, dotted curve is for $\nu=600$ and the dot-dashed curve is
for $\nu=900$.   
There is no constant effective viscosity curve that follows the N-Body
curve closely through the multi stream regions.
In regions with a given number of streams, the N-Body curve stays
around a curve for constant effective viscosity in the adhesion model. 
A remarkable fact is that the N-Body curve for the three stream region
at all the epochs follows the adhesion model curve for $\nu=\simeq 600$.   
Similar behaviour is seen for the five stream region which follows
$\nu \simeq 900$ though the range of scales and epochs over which this
can be resolved is somewhat limited.   

\begin{figure}
\includegraphics[width=3.3truein]{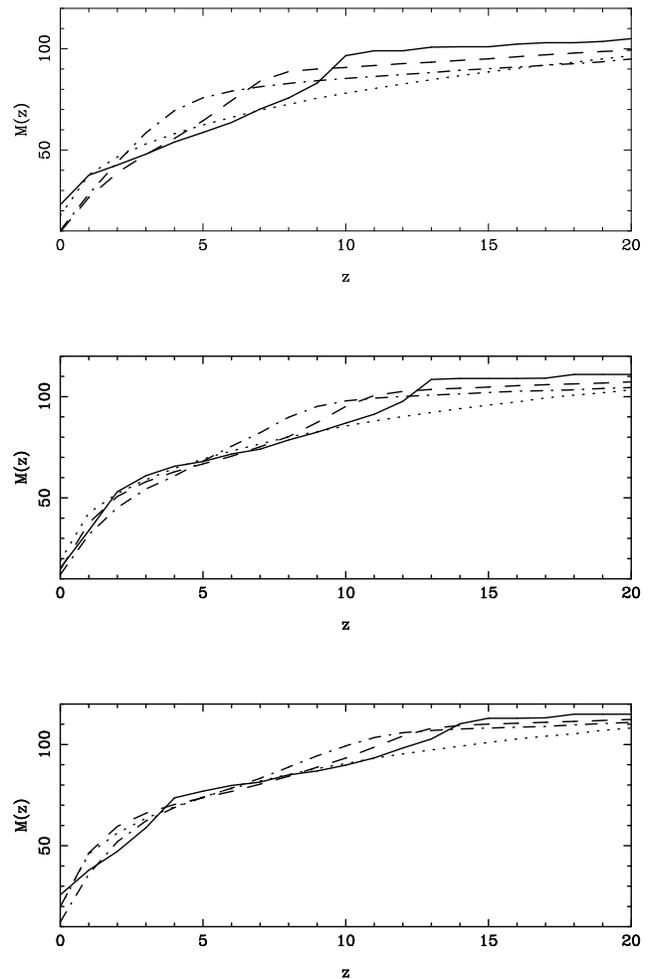}
\caption{This figure shows the mass enclosed within a distance $z$
  from the   centre of the multi stream region.  The top panel shows
  the curves for   $a=2$.  The solid curve is for N-Body simulation
  PM\_00L.  Other simulations are also plotted here T\_10L (dashed
  curve) and T\_40L (dot-dashed curve).   Dotted curve shows the mass
  enclosed in the one dimensional adhesion model with $\nu=600$.  The
  lower panel shows the same set of curves for $a=4$ and the middle
  panel is for $a=3$.} 
\label{den_adh_sub}
\end{figure}

Addition of substructure clearly changes the character of the problem
and the collapse is no longer one dimensional.  
However, the scale of the substructure is so small compared to the
wavelength of the plane wave that the large scale collapse is still
very close to planar.  
Figure~\ref{den_adh_sub} shows the mass enclosed within a distance $s$
from the centre of the multi stream region for simulations PM\_00L,
T\_10L and T\_40L. 
Density is averaged over all $x$ and $y$ for this plot in the same
manner as for figure~\ref{den_all}.  
Also plotted in the figure are curves for the adhesion model
($\nu=600$), where the calculation is done without taking substructure
into account.   
The motivation for such a comparison is to see the effect of
substructure on the favoured value of effective viscosity.
Substructure removes the sharp change in density at the boundaries of
$3$-stream, $5$-stream and $7$-stream regions and the curves for
T\_10L and T\_40L are smoother in the pancake region. 
The finite viscosity curve matches simulations with substructure
over a wider range of scales than with PM\_00L.  
There are no other noteworthy differences. 


\subsection{Mass Function}

Mass function of collapsed haloes in these simulations can be used to
understand the influence of plane wave collapse on substructure.   
Collapsed structures form in these simulations primarily due to
initial density fluctuations at small scales, with some modulation by
the collapse of the plane wave.  
In this section we study the effect of the collapsing plane wave on
the mass function of collapsed haloes.

These haloes were selected with the friends-of-friends (FOF) algorithm
using a linking length of $l=0.2$ grid length. 
The initial power spectrum has a peak at the scale corresponding to
$1/8$ of the simulation box, or $16$ grid lengths.  
Thus typical haloes will have a Lagrangian radius of about $8$ grid
lengths and should contain about $500$ particles.
Thus a cutoff of $50$ or more particles for haloes is reasonable for
this study. 

\begin{figure*}
\includegraphics[width=7truein]{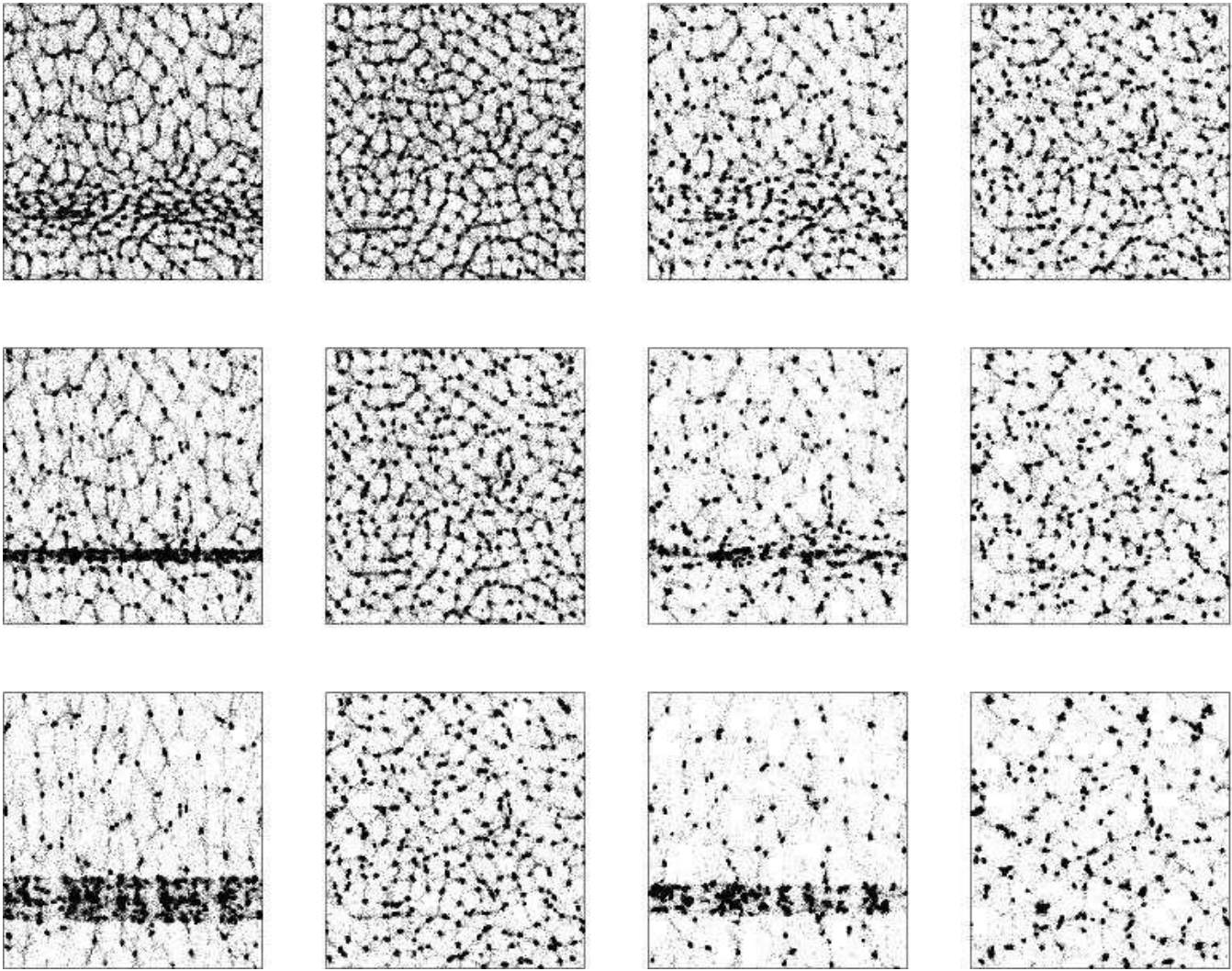}
\caption{This figure shows the effect of plane wave on evolution of small
  scale fluctuations.  Panels in this figure show a slice from N-Body
  simulations.  The top row is for $a=0.5$, middle row is for $a=1$ and the
  lower row is for $a=2$.  The left column is for T\_05L, the second
  column is for T\_05, the third column is for T\_20L and the right column is
  for T\_20.}
\label{slices}
\end{figure*}

\begin{figure}
\includegraphics[height=3.3truein,angle=270]{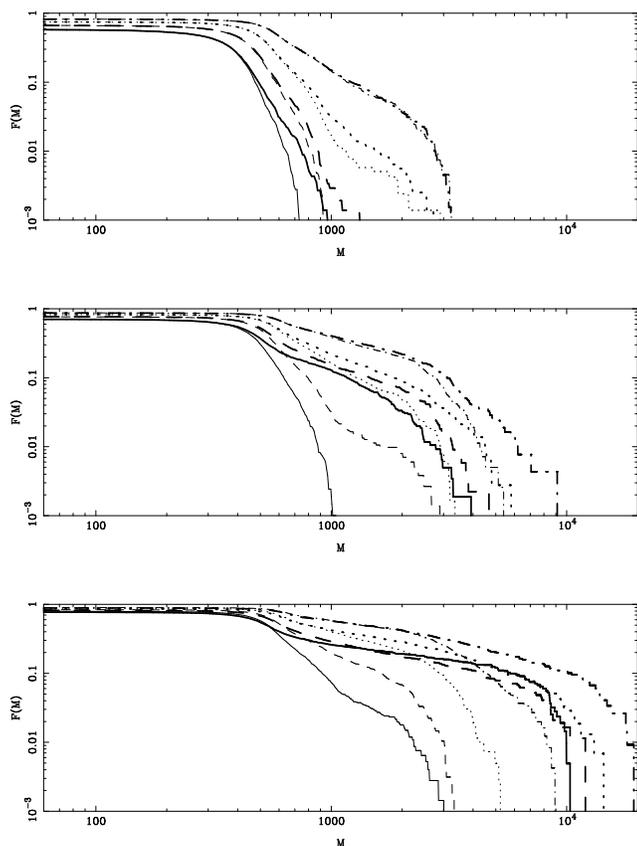}
\caption{This figure shows the cumulative mass function $F(M)$ as a function 
  of mass $M$ in N-Body simulations.  The top panel is for $a=0.5$, the middle
  panel is for $a=1$ and the lower panel is for $a=2$.  Curves are shown for
  T\_05 (solid curve), T\_05L (thick solid curve), T\_10 (dashed curve), T\_10L
  (thick dashed curve), T\_20 (dotted curve), T\_20L (thick dotted curve),
  T\_40 (dot-dashed curve) and T\_40L (thick dot-dashed curve).}
\label{massf}
\end{figure}

In absence of the plane wave, the only perturbations are at small
scales.   
The small scale perturbations are concentrated around a given mass
scale and the mass function is also peaked around this mass at early
epochs.   
At late epochs mergers lead to formation of haloes with a larger mass
and the range of masses is greater for models with a larger amplitude
of fluctuations. 
Figure~\ref{slices} shows these features in the distribution of
particles. 
These features can also be seen in figure~\ref{massf} where mass
fraction $F(M)$ for $a=0.5$, $1.0$ and $2.0$ is plotted in different panels.   
$F(M)$ is the fraction of total mass in collapsed haloes with halo mass above
$M$. 

Adding the plane wave at a much larger scale than the small scale
fluctuations essentially pushes much of the mass into the pancake
region, leaving a small fraction of matter in the under dense regions
that occupy much of the volume. 
Growth of small scale fluctuations in the under dense regions is
inhibited whereas growth of fluctuations in the pancake region is
enhanced, this is seen clearly in the slices from simulations shown in
figure~\ref{slices}.   
Higher background density in the pancake region leads to rapid growth
of perturbations, mergers of haloes also lead to formation of massive
clumps.  
These effects become more pronounced at late epochs and result in a
shift of mass function towards larger masses, indeed haloes at two
distinct mass scales are present.
Low mass clumps in under dense regions have the mass expected of
haloes in regions where small scale fluctuation dominate whereas
haloes of a much higher mass are present in the pancake region.  
Figure~\ref{massf} shows these two mass scales very clearly.  

Total mass in collapsed haloes does not change significantly with the 
addition of the plane wave.
Indeed for simulations T\_40L and T\_40, mass function is the same
at $a=0.5$ as small scales dominate. 
At late times ($a=2$), the effect of plane wave makes the mass
function of T\_05L, T\_10L and T\_20L similar.  

Not surprisingly, presence of large scale power leads to formation of
more massive haloes.
However it does not seem to enhance the total mass in collapsed haloes.


\section{Discussion}

In this paper we studied the effect of substructure on collapse of a
plane wave.  
The key conclusions of the present study of the role of substructure
are: 
\begin{itemize}
\item
The pancake formed due to collapse of the plane wave is thinner if the 
in falling material is formed of collapsed substructure. 
\item
We show that collisions between clumps lead to enhancement of
velocities transverse to the direction of large scale collapse.  
\item
We show that in simulations with substructure where collisions are suppressed,
pancakes are not thinner. 
\item
Thus collision induced enhancement of motions transverse to the collapsing
plane wave takes away kinetic energy from the direction of infall and leads to
thinner pancakes.  
\item
Presence of large scale power shifts the mass function towards larger
masses. 
There is, however, no change in the total mass in collapsed haloes. 
\end{itemize}

The points outlined above essentially relate to coupling of density
fluctuations at well separated scales.  
Each of these points refers to a measurable effect of such a
coupling. 
The nature of large scale fluctuation, a single plane wave, does not
allow us to estimate the effect in terms of statistical indicators
like the power spectrum. 
We plan to study these aspects with larger ($256^3$) simulations where
the large scale collapse will also be generic.
Large, high resolution studies are needed as $128^3$ simulations with
particle mesh code have not shown any large effect in power spectrum
at late times \cite{trpwr}. 

Another important point to consider is that we have considered two
well separated scales for fluctuations and there is no infall once
fluctuations at the larger scales collapse. 
Numerical experiments that can shed light on effects of this feature
are also required to improve our understanding of issues.

We also compared the collapse of a plane wave in an N-Body with the 
collapse in the adhesion model with a finite effective viscosity.
We found that:
\begin{itemize}
\item
The adhesion model predicts the variation of density very well with a
constant effective viscosity in regions with a given number of streams.
\item
Regions with a given number of streams coincide with the adhesion
model with the same value of effective viscosity at all epochs.
\end{itemize}


\section*{Acknowledgements}

JSB thanks R.Nityananda, T.Padmanabhan and K.Subramanian for useful
discussions and suggestions.  
JSB also thanks Varun Sahni and Uriel Frisch for a useful discussion on
related issues. 
Numerical experiments for this study were carried out at cluster computing
facility in the Harish-Chandra Research Institute
(http://cluster.mri.ernet.in). 
This research has made use of NASA's Astrophysics Data System. 


\label{lastpage}

\end{document}